\documentstyle[preprint,aps,epsf]{revtex}
\tighten
\begin{document}
\draft
\title{Quantum mechanics on moduli spaces.}
\author{Ian G. Moss and Noriko Shiiki}
\address{
Department of Physics, University of Newcastle Upon Tyne, NE1 7RU U.K.
}
\date{June 1998}
\maketitle
\begin{abstract}
It has been assumed that it is possible to approximate the interactions
of quantized BPS solitons by quantising a dynamical system induced on a
moduli space of soliton parameters. General properties of the reduction
of quantum systems by a Born-Oppenheimer approximation are described
here and applied to sigma models and their moduli spaces in order to
learn more about this approximation. New terms arise from the reduction
proceedure, some of them geometrical and some of them dynamical in
nature. The results are generalised to supersymmetric sigma models,
where most of the extra terms vanish.
\end{abstract}
\narrowtext
\section{INTRODUCTION}

In most physical problems, the degrees of freedom that are of interest
are only a small fraction of the total number of degrees of freedom
present. Some kind of reduction of the system has been performed. The
way in which this reduction proceeds from quantum field theory to a
moduli space of topological solitons is the subject of this paper.

Soliton solutions can give rise to particle states \cite{coleman} and
it has been predicted that their low-energy interations can be reduced
to a simplified quantum mechanical problem in a set of collective
coordinates \cite{gibbons,gauntlett}. One way in which this can be used
is to indicate a duality between two quantum theories, where the
soliton particle states in one theory become the fundamental particles
in an equivalent quantum theory. A particular example can be seen in
the monopole bound states which are important for the duality between
the large and small coupling constant limits of the heterotic string
theory \cite{sen}.

In a range of different models there are classical multiple-soliton
solutions that saturate Bogomol'nyi bounds and are parameterised by
moduli. When the solitons move slowly and interact they deform
adiabatically and the moduli trace out a path in a moduli space. The
equations governing the path are equivalent to the equations for a
point particle moving on a curved space \cite{manton,ward}.

We would expect to be able to describe the interactions of quantized
solitons by quantising the dynamical system induced on the moduli
space. We will call this proceedure {\it truncation}. However, the
quantisation rules of the truncated theory are often ambiguous without
additional assumptions. If we impose general covariance, reflecting the
freedom to choose coordinates on the moduli space, then the Hamiltonian
operator takes the form
\begin{equation}
-\case1/2\Delta+\xi R+\dots\label{ambig}
\end{equation}
where $\Delta$ is the Laplacian on the moduli space and ambiguities are
reduced to the coefficients $\xi$ of the Ricci scalar $R$ and other
curvature invariants \cite{dewitt}. These ambiguities can be further
reduced by introducing supersymmetry, in which case it has been argued
that $\xi=0$ \cite{witten,alfaro}.

In this paper we describe the alternative to truncating the full
theory, which is to {\it reduce} it by a Born-Oppenheimer
approximation. In a purely bosonic theory we will see that general
covariance breaks down in the sense that a potential arises on the
moduli space after integrating out the fluctuations that are orthogonal
to the moduli space.

In the basic situation we begin with a classical dynamical system in
flat space which has a potential $V$ with a set of degenerate minima
${\cal M}$. We find that the original quantum theory reduces to a
quantum theory on ${\cal M}$ with Hamiltonian
\begin{equation}
H_R=-\case1/2\Delta_{\cal M}+U
+\case1/4R_{\cal M}-\case1/8|{\bf k}|^2+
+\dots
\end{equation}
The two purely geometrical terms depend on the intrinsic curvature
scalar $R_{\cal M}$ and the (traced) extrinsic curvature ${\bf k}^I$.
These geometrical terms where first written down by Maraner
\cite{maraner}. The dynamical terms denoted by $U$ depend upon
derivatives of the potential. Those depending on up to two extrinsic
derivatives of the potential can be written
\begin{equation}
U=\case1/2{\rm tr}({\bf \omega})+\case1/{16}{\rm tr}
\left({\bf \omega}^{-1}{}^\parallel\nabla{\bf \omega}\cdot
{\bf \omega}^{-1}{}^\parallel\nabla{\bf \omega}\right)\label{vacuum}
\end{equation}
where $\omega$ is a matrix, ${\bf\omega}^2$ is the hessian matrix of
$V$ and the covariant derivative is taken tangentially to ${\cal M}$.

The application of these results to sigma models is considered in
section 3. In these models the original configuration space, consisting
of maps from physical space to a curved target space, is both infinite
and curved. The infinities can be dealt with by regularisation. The
curvature would lead in general to ambiguities even before the
reduction. However, as shown in appendix A, these ambiguities are not
present if the original space is a Riemannian symmetric space. Most
models of interest fall into this class.

The results are generalised to supersymmetric sigma models in section
4. The point at issue is whether the reduction leads to a
superpotential on the moduli space and alters the hamiltonian operator.
We find that, to leading order, the theory reduces to the Hodge
Laplacian acting on $p-forms$,
\begin{equation}
H_R=-\case1/2\Delta_{\cal M}+\dots
\end{equation}
confirming previous expectations \cite{witten,alfaro,gauntlett}.
Possible developments of this work are addressed in the conclusion,
where we also discuss the advantages of the Hamiltonian approach over a
Feynmann diagram approach.

\section{BORN-OPPENHEIMER REDUCTION}

Consider a classical dynamical system in flat space which has a
degenerate set of stable points ${\cal M}$. If the potential is
sufficiently steep, then energy spectrum of the quantum system
typically falls into seperate bands. An adiabatic approximation scheme
can then be used to reduce the quantum system to a quantum system on
${\cal M}$.

We shall retrict attention to the case where the potential has an
analytic expansion about its minima. We take a set of generalised
coordinates $(x^a,x^I)$ such that $x^I=0$ on ${\cal M}$. In the lowest
energy bands, the value of $x^I$ fluctuates over a narrow range of
values determined by the eigenvalues of a matrix $\bf \omega$, where
\begin{equation}
{\bf \omega}^2=\partial_I\partial_JV.
\end{equation}
The expansion parameter of the approximation scheme is given by the
width of an energy band divided by the smallest eigenvalue.

The first step is to rewrite the Hamiltonian operator,
\begin{equation}
H=-\case1/2\Delta+V
\end{equation}
in the new coordinate system. We choose normal coordinates $x^I$ along
the principal directions of $\partial_I\partial_JV$. It is always
possible to find two mutually orthogonal sets of derivatives
$\hat\partial_a$ and $\partial_I$, where
\begin{equation}
\hat\partial_a=\partial_a-N^I{}_a\partial_I.
\end{equation}
This implies that the metric can be written
\begin{equation}
{\bf g}=\gamma_{ab}dx^adx^b+\gamma_{IJ}
(dx^I+N^I{}_adx^a)(dx^J+N^J{}_bdx^b).
\end{equation}
The metric components have normal coordinate expansions
\begin{eqnarray}
\gamma_{IJ}&=&\delta_{IJ}\label{nca}\\
N^I{}_a&=&-a_J{}^I{}_ax^J\\
\gamma_{ab}&=&\sigma_{ab}+2k_{I\,ab}x^I+k_{I\,ac}k_{J\,b}{}^cx^Ix^J\label{nce}
\end{eqnarray}
where the extrinsic curvature ${\bf k}^I$ and the torsion ${\bf
a}_I{}^J$ are defined in appendix B. These expansions are exact if the
space is flat, but we will keep in mind the possibility of generalising
these results to curved spaces.

In the new coordinate system, the Hamiltonian becomes
\begin{equation}
H=\case1/2|g|^{-1/2}\hat\partial_a|g|^{1/2}\gamma^{ab}\hat\partial_b+
\case1/2|g|^{-1/2}\partial_I|g|^{1/2}\delta^{IJ}\partial_J+V
\end{equation}
We regard the $x^I$ as small and expand the Hamiltonian as a series of
terms $H=H_0+H_1+\dots$. The potential has a series expansion
\begin{equation}
V=\case1/2(\omega^2)_{IJ}x^Ix^J+\case1/6V_{IJK}x^Ix^Jx^K+\dots\label{pot}
\end{equation}
where $\omega_{IJ}$ is a diagonal matrix with eigenvalues $\omega_I$.
The first few terms in the full Hamiltonian are then
\begin{eqnarray}
H_0&=&-\case1/2\delta^{IJ}\partial_I\partial_J+
\case1/2(\omega^2)_{IJ}x^Ix^J\label{unpert}\\
H_1&=&-\case1/2\,{\rm tr}({\bf k}^I)\partial_I+\case1/6V_{IJK}x^Ix^Jx^K
\\
H_2&=&-\case1/2{\cal D}\cdot{\cal D}
+\case1/2{\rm tr}({\bf k}_I{\bf k}^J)x^I\partial_J+
\case1/{24}V_{IJKL}x^Ix^Jx^Kx^L\label{htwo}
\end{eqnarray}
The dot denotes a scalar product using the intrinsic metric on ${\cal
M}$. We have introduced a covariant derivative ${\cal D}_a$, where
${\cal D}_a\sigma_{ab}=0$ and
\begin{equation}
{\cal D}_af=(\partial_a+a_J{}^I{}_ax^J\partial_J)f\label{dcal}
\end{equation}
for scalar functions $f$.

The next step is to apply degenerate perturbation theory. We take the
unperturbed harmonic oscillator Hamiltonian $H_0$ together with a set
of states $f_n$, where $H_0f_n=E_nf_n$. The perturbation theory is
described in appendix C and gives a reduced Hamiltonian operator $H_R$
acting on the space of wave functions $\psi(x^a)$,
\begin{equation}
H_R=\langle H_0\rangle_{00}+\langle H_1\rangle_{00}+\langle
H_2\rangle_{00}+
\sum_{m\ne0}\langle H_1\rangle_{0m}\left(E_0-E_m\right)^{-1}\langle
H_1\rangle_{m0}+O(\omega^{-1})\label{degen}
\end{equation}
The matrix elements defined by
\begin{equation}
\langle H\rangle_{nm}=\int f_n^*\,H\,f_m \prod_Idx^I\label{matrix}
\end{equation}
are allowed to be operator-valued.

The first term $\langle H_0\rangle_{00}$ consists of the vacuum energy
of the system of unperturbed harmonic oscillators (\ref{unpert}),
\begin{equation}
E_0=\case1/2{\rm tr}({\bf \omega}).
\end{equation}
The second term $\langle H_1\rangle_{00}$ vanishes because the operator
is odd under the inversion $x^I\to-x^I$. For the next term $\langle
H_2\rangle_{00}$, with $H_2$ given by equation (\ref{htwo}), we set
\begin{equation}
\langle {\cal D}_a\rangle_{mn}=
\delta_{mn}{}^\parallel\nabla_a+\langle {\cal A}_a\rangle_{mn}
\end{equation}
where ${}^\parallel\nabla_a$ is defined in appendix B and ${\cal A}$
arises form the action of ${\cal D}_a$ on the states $f_n$ in equation
(\ref{matrix}). The matrix element of the derivative of an energy
eigenstate is given by a standard identity,
\begin{equation}
\langle {\cal A}_a\rangle_{mn}={([{\cal D}_a,H_0])_{mn}\over E_n-E_m}
\end{equation}
After using equation (\ref{dcal}) for ${\cal D}$, and comparing with
equation (\ref{grado}), the commutator produces a covariant derivative,
\begin{equation}
\langle {\cal A}_a\rangle_{mn}=\cases{
\case1/2{}^\parallel\nabla_a\omega_{IJ}\,\langle
x^Ix^J\rangle_{mn}&$E_m<E_n$\cr
0&$m=n$\cr}.
\end{equation}
We can now evaluate
\begin{equation}
\langle {\cal D}\cdot{\cal D}\rangle_{00}=\Delta_{\cal M}-
\case1/8{\rm tr}\left({\bf \omega}^{-1}{}^\parallel\nabla{\bf
\omega}\cdot
{\bf \omega}^{-1}{}^\parallel\nabla{\bf \omega}\right)
\end{equation}
where the covariant Laplacian $\Delta_{\cal M}=
{}^\parallel\nabla\cdot{}^\parallel\nabla$. We also have, using
(\ref{gauss}),
\begin{equation}
\langle{\rm tr}({\bf k}_I{\bf k}^J)x^I\partial_J\rangle_{00}=
\case1/2{\rm tr}({\bf k}_I{\bf k}^I)=
\case1/2|{\bf k}|^2-\case1/2R_{\cal M}
\end{equation}
where $|{\bf k}|^2={\rm tr}({\bf k}^I){\rm tr}({\bf k}_I)$.
Therefore we get
\begin{equation}
\langle H_2\rangle_{00}=-\case1/2\Delta_{\cal M}-\case1/4|{\bf k}|^2
+\case1/4R_{\cal M}+\case1/{16}
{\rm tr}\left({\bf \omega}^{-1}{}^\parallel\nabla{\bf \omega}\cdot
{\bf \omega}^{-1}{}^\parallel\nabla{\bf \omega}\right)
+\case1/{32}\sum_{IJ}\omega_I^{-1}\omega_J^{-1}V_{IJIJ}
\end{equation}
The last term given in equation (\ref{degen}) leads to
\begin{equation}
\case1/8\,|{\bf k}|^2
-\case1/{48}\sum_{IJK}(\omega_I+\omega_J+\omega_K)^{-1}
\omega_I^{-1}\omega_J^{-1}\omega_K^{-1}(V_{IJK})^2
-\case1/{32}\sum_{IJK}\omega_I^{-2}\omega_J^{-1}\omega_K^{-1}V_{IJJ}V_{IKK}
\end{equation}

Putting the terms together gives a final expression for the reduced
Hamiltonian,
\begin{equation}
H_R=-\case1/2\Delta_{\cal M}+U
+\case1/4R_{\cal M}-\case1/8|{\bf k}|^2+O(\omega^{-1})\label{result}
\end{equation}
The terms are divided into purely geometrical terms and dynamical terms
which are collected into $U$, where
\begin{eqnarray}
U&=&\case1/2{\rm tr}({\bf \omega})+\case1/{16}{\rm tr}
\left({\bf \omega}^{-1}{}^\parallel\nabla{\bf \omega}\cdot
{\bf \omega}^{-1}{}^\parallel\nabla{\bf \omega}\right)
+\case1/{32}\sum_{IJ}\omega_I^{-1}\omega_J^{-1}V_{IJIJ}\nonumber\\
&&-\case1/{48}\sum_{IJK}(\omega_I+\omega_J+\omega_K)^{-1}
\omega_I^{-1}\omega_J^{-1}\omega_K^{-1}V_{IJK}{}^2
-\case1/{32}\sum_{IJK}\omega_I^{-2}\omega_J^{-1}\omega_K^{-1}V_{IJJ}V_{IKK}
\end{eqnarray}
The vacuum energy, which is of order $\omega$, is the dominant term in
this expression. The next term depends on the gradient of the normal
mode frequencies along the moduli space and includes the effects of the
twisting, or torsion, of the normal mode directions. The remaining
terms contain the effects of higher derivatives of the potential.

The geometrical terms where discovered in the restricted situation of a
surface embedded in $R^3$ by Jensen and Koppe \cite{jensen} and more
generally by Maraner \cite{maraner}. The covariant forms for the
dynamical terms beyond the obvious vacuum energy term are new, as far
as we know.

\section{BOSONIC FIELD THEORY}

The non-linear sigma-model in two space dimension $x^\mu$ and one time
dimension $t$ has fields $\phi^i$ taking values in a curved target
space ${\cal T}$. We will take the Lagrangian to be $L=T-V$, where
\begin{eqnarray}
T&=&{1\over 2}\int g_{ij}\partial_t\phi^i\partial_t\phi^j\,d^2x\\
V&=&{1\over 2}\int g_{ij}\partial_\mu\phi^i\partial_\mu\phi^j\,d^2x
\end{eqnarray}
Depending upon the topology of ${\cal T}$, the classical field
equations can have static solutions with finite energy, or topological
solitons. These solutions minimise the potential energy for a given
topological class.

When ${\cal T}$ is a compact Kahler manifold the soliton solutions can
be represented by holomorphic maps from the complex plane into ${\cal
T}$ \cite{ward,din,ruback}. They depend on a continuous set of
parameters, or moduli, which can be interpreted as the positions and
charges of individual lumps.

The initial situation is similar to the models considered in the
previous section, except that the configuration space is no longer
finite and flat. The introduction of curvature leads to the appearance
of curvature invariants in the initial Schr\"odinger equation. However,
we will take the target space to be a Riemannian symmetric space. The
space of fields inherits this symmetry and is also a Riemannian
symmetric space. In this case the curvature invariants are constants
which can be absorbed into an overall phase factor.

The soliton solutions $\phi_0$ will be parameterised by a set of
coordinates $x^a$ belonging to the moduli space. We shall expand the
Lagrangian about these solutions using standard background field
methods \cite{luis} and express the result in terms of time-dependent
coordinates $x^a$ and normal coordinates $x^I$. This will enable us to
use the results of the previous section.

The tangent vector $\xi$ to the geodesic from $\phi$ to $\phi_0$
provides a convenient measure of the displaced field. We also introduce
the vectors ${\bf e}_\mu$ with target space components
\begin{equation}
{\bf e}_\mu{}^i=\partial_\mu\phi^i
\end{equation}
These can be parallel transported back from the tangent space at $\phi$
to the tangent space at $\phi_0$ to produce a field ${\bf e}'_\mu$. The
tangent vector $\xi$ commutes with ${\bf e}_\mu$, and consequently
$\nabla_\xi{\bf e}_\mu=D_\mu\xi$, where $D_\mu$ is the gradient along
${\bf e}_\mu$. This relation simplifies the Taylor series expansion for
${\bf e}'$,
\begin{equation}
{\bf e}'_\mu={\bf e}_\mu+D_\mu\xi+\case1/2R(\xi,{\bf e}_\mu)\xi
+\case1/6R(\xi,D_\mu\xi)\xi+\dots\label{serie}
\end{equation}

We will define
\begin{equation}
({\bf u},{\bf v})=\int g_{ij}(\phi_0)u^iv^j\,d^2x\label{product}
\end{equation}
By parallel transport of the Lagrangian density we can write the
potential in the form
\begin{equation}
V=\case1/2({\bf e}'_\mu,{\bf e}'{}_\mu)
\end{equation}
The potential therefore has a series expansion in $\xi$ of the form
$V=V_0+V_1+V_2\dots$ where
\begin{eqnarray}
V_0&=&\case1/2({\bf e}_\mu,{\bf e}_\mu)\label{vzero}\\
V_1&=&({\bf e}_\mu,D_\mu\xi)\label{vone}\\
V_2&=&\case1/2(D_\mu\xi,D_\mu\xi)+
\case1/2({\bf e}_\mu,R(\xi,{\bf e}_\mu)\xi)\label{vtwo}\\
V_3&=&\case2/3(D_\mu\xi,R(\xi,{\bf e}_\mu)\xi)\label{vthree}\\
V_4&=&\case1/6(D_\mu\xi,R(\xi,D_\mu\xi)\xi)
+\case1/6(R(\xi,{\bf e}_\mu)\xi,R(\xi,{\bf e}_\mu)\xi)
\end{eqnarray}
Integration by parts in (\ref{vone}) produces the field equation
\begin{equation}
D_\mu{\bf e}_\mu=
\partial_\mu e_\mu{}^i-\Gamma^i{}_{jk}e_\mu{}^je_\mu{}^k=0\label{fe}
\end{equation}
Integration by parts on (\ref{vtwo}) produces the operator
\begin{equation}
\Delta_f\xi=-D_\mu D_\mu\xi-R(\xi,{\bf e}_\mu){\bf e}_\mu,\label{fluc}
\end{equation}
which describes fluctuations about the moduli space. In the usual
analysis of the sigma model, the inverse of this propagator would be
the Green function in the background field approximation.

The normalised positive modes of the fluctuation operator will be
denoted by ${\bf u}_I$ and their eigenvalues by $(\omega_I)^2$. The
fluctuation operator also has a set of zero modes ${\bf
u}_a=\partial_a\phi_0$. All of the modes are parameterised by the
collective coordinates $x^a$. We define the remaining coordinates in
terms of the displacement vector,
\begin{equation}
\xi=x^I{\bf u}_I\label{xex}
\end{equation}
The potential then has a series expansion in the normal coordinates
identical to equation (\ref{pot}) used in the previous section. From
equation (\ref{vthree}), \begin{equation}
V_{IJK}=4C_{(IJ)K}\label{pott}
\end{equation}
where
\begin{equation}
C_{IJK}=(D_\mu{\bf u}_I,R({\bf u}_J,{\bf e}_\mu){\bf u}_K).\label{cdef}
\end{equation}
The coefficients are symmetrical, since $C_{(IJ)K}=C_{(IJK)}$. They are
also trace-free if the zero modes $u_a$ are included,
$C^I{}_{IJ}+C^a{}_{aJ}=0$.

The kinetic energy has a similar series expansion. Since
\begin{equation}
T=\case1/2({\bf e}'_t,{\bf e}'{}_t),
\end{equation}
we need only replace ${\bf e}_\mu$ by ${\bf e}_t$ and $D_\mu\xi$ by
$D_t\xi$ in equations (\ref{vzero}-\ref{vthree}). However, both the
$x^I$ and the $x^a$ depend on time, so that the time derivative of
equation (\ref{xex}) implies
\begin{equation}
D_t\xi=\dot x^I\,{\bf u}_I+\dot x^a\,x^ID_a{\bf u}_I\label{dtxi}
\end{equation}
where the covariant derivative
\begin{equation}
D_a u^i=\partial_a u^i+u_a{}^k\Gamma^i{}_{jk}u^j.
\end{equation}
Also, since ${\bf u}_a=\partial_a\phi_0$,
\begin{equation}
{\bf e}_t=\partial_t\phi_0=\dot x^a{\bf u}_a.\label{et}
\end{equation}
Equations (\ref{dtxi}) and (\ref{et}) allow terms in the series
expansion of the kinetic energy to be grouped as a quadratic polynomial
in the generalised velocities,
\begin{equation}
T=\case1/2\gamma_{ab}\dot x^a\dot x^b+\case1/2\gamma_{IJ}
(\dot x^I+N^I{}_a\dot x^a)(\dot x^J+N^J{}_b\dot x^b)
\end{equation}
To second order in $x^I$,
\begin{eqnarray}
\gamma_{IJ}&=&\delta_{IJ}-\case1/3R_{IKJL}x^Kx^L\\
N^I{}_a&=&-a_J{}^I{}_ax^J\\
\gamma_{ab}&=&\sigma_{ab}+2k_{I\,ab}x^I+k_{I\,ac}k_{J\,b}{}^cx^Ix^J
-R_{aIbJ}x^Ix^J
\end{eqnarray}
The coefficients can be obtained by examining equations
(\ref{vzero}-\ref{vthree}),
\begin{eqnarray}
\sigma_{ab}&=&({\bf u}_a,{\bf u}_b)\\
k^I{}_{ab}&=&({\bf u}_a, D_b{\bf u}_I)\\
a_I{}^J{}_a&=&({\bf u}_I, D_a{\bf u}_J)\\
R_{IJKL}&=&({\bf u}_I,R({\bf u}_K,{\bf u}_L){\bf u}_J)
\end{eqnarray}
These results recover the normal coordinate expansion of the previous
section (\ref{nca})-(\ref{nce}), with extra terms due to the curvature
of the original configuration space. We now have explicit expressions
for the curvature and the extrinsic geometry of the space of soliton
parameters in terms of eigenstates of the fluctuation operator.

The results of the previous section still apply, except for the
addition of terms depending on the target space curvature. A similar
analysis to the previous section shows that two extra terms
\begin{equation}
\case1/4\sigma^{ac}\sigma^{bd}R_{abcd}+
\case1/6\sum_{IJ}\omega_I\omega_J^{-1}R_{IJIJ}
\end{equation}
should be included in the potential $U$.

Expressions involving derivatives of the eigenmodes can be rewritten
using the identity
\begin{equation}
({\bf u}_I,D_a{\bf u}_J)={({\bf u}_I,[D_a,\Delta_f]{\bf u}_J)\over
\omega_J^2-\omega_I^2}
\end{equation}
With the fluctuation operator $\Delta_f$ given explicitly in equation
(\ref{fluc}), the commutator becomes
\begin{equation}
({\bf u}_I,[D_a,\Delta_f]{\bf u}_J)=4C_{(IJ)a}
\end{equation}
where $C_{IJa}$ is defined as in equation (\ref{cdef}). This gives
\begin{eqnarray}
k_{I\,ab}&=&-{4\over\omega_I^2}C_{(Ia)b}\label{kexp}\\
{}^\parallel\nabla_a\omega_{IJ}&=&{4\over\omega_J+\omega_I}
C_{(IJ)a}\label{goexp}
\end{eqnarray}
where equation (\ref{grado}) has been used.

The dominant term in the reduced Hamiltonian should be the zero point
energy $\case1/2{\rm tr}(\omega)$. In general, in order to evaluate
this term it will be necessary to solve the eigenvalue problem for the
fluctuation operator numerically. Once this has been done, equations
(\ref{pott}), (\ref{kexp}) and (\ref{goexp}) could then be used to
evaluate the geometrical terms in the reduced Hamiltonian
(\ref{result}).

\section{SUPERSYMMETRIC REDUCTION}

We now turn to the reduction of an $N=2$ supersymmetric quantum system
defined on a Riemannian symmetric space. As before, we supose that the
system has a potential $V$ with a degenerate set of minima. The
existence of two supersymmetries requires that the space is a Kahler
manifold with a covariantly preserved complex structure $J$
\cite{zumino,alvarez}. We shall suppose that the moduli space of stable
points is also a Kahler manifold.

The variables consist of coordinates $x^\alpha$ and single complex
component fermions $\psi^\alpha$. We also assume the existence of an
operator $\pi_\alpha$, related to the momentum, and the following
commutation relations
\begin{eqnarray}
\{\psi^\alpha,\psi^{*\beta}\}&=&g^{\alpha\beta}\\
 \lbrack\pi_\alpha,x^\beta\rbrack&=&-i\delta_\alpha{}^\beta\\
 \lbrack\pi_\alpha,\psi^\beta\rbrack
&=&-i\Gamma^\beta{}_{\gamma\alpha}\psi^\gamma\\
 \lbrack\pi_\alpha,\pi_\beta\rbrack&=&
-iR_{\alpha\beta\gamma\delta}\psi^{*\gamma}\psi^\delta\label{coms}
\end{eqnarray}
involving the inverse metric $g^{\alpha\beta}$, connection components
$\Gamma^\beta{}_{\gamma\alpha}$ and curvature components
$R_{\alpha\beta\gamma\delta}$.

It is possible to represent the commutator relations by covariant
differential operators on a Hilbert space with the basis
\begin{equation}
\Psi(x)_{\alpha_1\dots\alpha_p}
\psi^{*\alpha_1}\dots\psi^{*\alpha_p}|0\rangle,
\end{equation}
where $\psi^\alpha|0\rangle=0$. This gives a representation in which
\begin{equation}
\pi_\alpha=-i(\partial_\alpha-
\Gamma_{\gamma\beta\alpha}\psi^{*\beta}\psi^\gamma)\label{pi}
\end{equation}
The action of $\pi$ is equivalent to $-i\nabla_\alpha$ acting on the
$p$-forms $\Psi$, where $\nabla_\alpha$ is the metric connection.

We shall assume that the classical system has two supersymmetries
generated by the supercharges
\begin{eqnarray}
Q^{(1)}&=&\psi^\alpha\pi_\alpha-\psi^{*\alpha}W_\alpha(x)\\
Q^{(2)}&=&(J\psi)^\alpha\pi_\alpha-(J\psi^*)^\alpha W_\alpha(x)
\end{eqnarray}
The reason for choosing this particular form for these supercharges
will become apparent when we come to discuss sigma models in the next
section. We will only require the combination
\begin{equation}
Q=\case1/2(Q^{(1)}+iQ^{(2)})=
\lambda^\alpha\pi_\alpha-\nu^\alpha W_\alpha(x)\label{sq}
\end{equation}
where the fermion fields
\begin{eqnarray}
\lambda&=&\case1/2(1+iJ)\psi\\
\nu&=&\case1/2(1+iJ)\psi^*
\end{eqnarray}
are purely holomorphic.

The Hamiltonian is given (up to a constant) by $H=\{Q,Q^*\}$, which can
be evaluated using the commutation relations (\ref{coms}),
\begin{eqnarray}
H&=&-\case1/2\nabla^2+g^{\alpha\beta}W_\alpha W^*_\beta
+iW^*_{\alpha;\beta}\lambda^\beta\nu^{*\alpha}
+iW_{\alpha;\beta}\lambda^{*\beta}\nu^\alpha\nonumber\\
&&-\case1/4R_{\alpha\beta\gamma\delta}
[\lambda^\alpha,\lambda^{*\beta}][\lambda^{*\gamma},\lambda^\delta]
-\case1/4R_{\alpha\beta\gamma\delta}
[\lambda^\alpha,\lambda^{*\beta}][\nu^\gamma,\nu^{*\delta}]\label{ha}
\end{eqnarray}
The moduli space lies at the minimum of the bosonic potential and is
parameterised by a set of complex coordinates $z^a$. We asume that the
superpotential $W_\alpha$ has a series expansion in normal coordinates
$z^I$,
\begin{equation}
W_{\alpha}=W_{\alpha\bar J}z^{\bar J}
+W_{\alpha\bar J\bar K}z^{\bar J}z^{\bar K}
+W_{\alpha\bar JK\bar L}z^{\bar J}z^Kz^{\bar L}+\dots\label{sup}
\end{equation}
The normal coordinates can be chosen to ensure that the boson mass
matrix is diagonal,
\begin{equation}
g^{\alpha\beta}W_{\alpha\bar J}W^*_{\beta K}=(\omega^2)_{\bar J
K}\label{bmass}
\end{equation}
For the fermion mass matrix $W_{\alpha;\beta}$, we need two seperate
sets of basis vectors, in general, in order to put it into diagonal
form. We can always choose one basis $({\bf e}_a,{\bf e}_I)$ to
coincide with the coordinate basis at $z^I=0$ for the fields $\lambda$.
Another basis $({\bf e}_{a'},{\bf e}_{I'})$ then has to be used for the
fields $\nu$. (This proceedure is similar to the independent unitary
rotations of the left and right chirality fermions in the standard
model). With the choice
\begin{equation}
{\bf e}_{I'}=\omega_I^{-1}W^\alpha{}_I{\bf e}_\alpha,
\end{equation}
the fermion mass matrix becomes diagonal,
\begin{equation}
W_{I'\bar J}=\omega_{I\bar J}.
\end{equation}
The fermion fields $\lambda^a$ and $\nu^{a'}$ are massless and remain
in the reduced quantum system. However, there is an important
difference between the finite and infinite dimensional situations. In
the infinite dimensional situation the basis ${\bf e}_{I'}$ may be
complete (in which case there are no massless $\nu^{a'}$ fermions). We
shall proceed assuming this to be the case.

Previously, we used a series expansion for the Hamiltonian to arrive at
the reduced theory. However, as shown in appendix C, for a
supersymmetric theory we need only consider the supercharge. The
supercharge (\ref{sq}) depends on the normal coordinates through the
superpotential and through the connection coefficients in $\pi_\alpha$.
In the basis ${\bf e}_\alpha=({\bf e}_a,{\bf e}_I)$,
\begin{equation}
\Gamma_{\alpha\beta a}\psi^{*\alpha}\psi^\beta\sim
\Gamma_{b\bar ca}\lambda^{\bar c}\lambda^b
+k_{\bar Iab}\lambda^{\bar I}\lambda^a
+a_{I\bar Ja}\lambda^{\bar J}\lambda^I
+a'{}_{\bar I'J'a}\nu^{J'}\nu^{\bar I'}\label{cterms}
\end{equation}
in the limit $z^I=0$.

The first few terms in the series expansion of the supercharge
(\ref{sq}) are now
\begin{eqnarray}
Q_0&=&-i\lambda^I\partial_I-\nu^{I'}W_{I'\bar J}z^{\bar J}\\
Q_1&=&-i\lambda^a{\cal D}_a-\nu^{I'}W_{I'\bar J\bar K}z^{\bar J}z^{\bar
K}\\
Q_2&=&-i\lambda^\alpha R_{\beta\gamma\bar I\alpha}z^{\bar I}
(\lambda^{*\gamma}\lambda^\beta-\nu^\gamma\nu^{*\beta})+
\nu^{I'}W_{I'\bar J\bar K\bar L}z^{\bar J}z^{\bar K}z^{\bar L}
\end{eqnarray}
where ${\cal D}_a$ now includes the connection terms (\ref{cterms}).

The full quantum theory can now be reduced to the moduli space by
following the methods described in appendix C. We define the fermion
vacuum state $|0_T\rangle$ which is anihilated by two sets of
anihilation operators
\begin{equation}
\lambda_-^I=\case1/{\sqrt{2}}(\lambda^I+i\nu^{I'}),\quad
\lambda_-^{\bar I}=\case1/{\sqrt{2}}
(\lambda^{\bar I}+i\nu^{\bar I'})
\end{equation}
As in section 2, we build up from $|0_T\rangle$ a set of oscillator
states $f_n$ which satisfy $H_0f_n=E_nf_n$, where $H_0$ is the
unperturbed Hamiltonian. In the supersymmetric case, $E_0=0$.

The reduced Hamiltonian is given in terms of the reduced supercharge by
\begin{equation}
H_R=\{Q_R,Q_R^*\}
\end{equation}
where
\begin{equation}
Q_R=\langle Q_1\rangle_{00}+
\sum_{n\ne0}\langle Q_1\rangle_{0n}(E_0-E_n)^{-1}
\langle H_1\rangle_{n0}+O(\omega^{-1})
\end{equation}
The covariant derivatives are treated as in section 2, that is we write
\begin{equation}
\langle{\cal D}_a\rangle_{0n}=\delta_{0n}{}^\parallel\nabla_a
+\langle{\cal A}_a\rangle_{0n}
\end{equation}
In the supersymmetric case we include the fermion terms from equation
(\ref{cterms}),
\begin{equation}
\langle{\cal A}_a\rangle_{0n}=
{}^\parallel\nabla_a\omega_{\bar IJ}\langle z^{\bar I}z^J\rangle_{0n}
-k_{\bar Iba}\langle \lambda^{\bar I}\lambda^b\rangle_{0n}
-a_{I\bar Ja}\langle \lambda^{\bar J}\lambda^I\rangle_{0n}
+a'{}_{I'\bar J'a}\langle \nu^{I'}\nu^{\bar J'}\rangle_{0n}
\end{equation}
At leading order we recover the same result that we would obtain by a
trivial truncation of the theory,
\begin{equation}
Q_R=-i\lambda^a{}^\parallel\nabla_a+O(\omega^{-1})
\end{equation}
As discussed in appendix A, the reduced states can be identified with
antiholomorphic forms on the moduli space and the the reduced
Hamiltonian can be identified with the Laplacian. At this order the
reduced theory is identical with the truncated theory obtained under
the simplest assumptions \cite{gauntlett}.

\section{SUPERSYMMETRIC SIGMA MODELS}

The supersymmetric sigma model in two space and one time dimension has
Lagrangian
\begin{equation}
L={1\over 2}\int\left(-g_{ij}\partial^\mu\phi^i\partial_\mu\phi^j
+ig_{ij}\bar\chi^i\gamma^\mu D_\mu\chi^j\right)d^2x
\end{equation}
where $\mu=0,1,2$, and $\chi$ is a two-component majorana spinor. If
the target space is a K\"ahler manifold, with complex structure $J$,
then the model has two supersymmetries. We shall consider the reduction
of this theory.

It is convenient to use a complex representation for the fermion
fields, with
\begin{equation}
\chi=\pmatrix{\psi\cr\psi^*},
\end{equation}
and a complex spatial coordinate $z=x+iy$. There are two complex
supercharges in this representation,
\begin{eqnarray}
Q^{(1)}&=&\int\left(g_{ij}\psi^i\partial_t\phi^j
-2g_{ij}\psi^{*i}\partial_z\phi^j\right)d^2z\\
Q^{(2)}&=&\int\left(g_{ij}(J\psi)^i\partial_t\phi^j
-2g_{ij}(J\psi)^{*i}\partial_z\phi^j\right)d^2z
\end{eqnarray}
These can be used to generate the classical supersymmetry
transformations by transforming to phase space and using Dirac
brackets. The supercharges produce the Hamiltonian by Dirac brackets
\begin{equation}
H=\case{i}/2\{Q^{(1)},Q^{(1)*}\}_{\rm DB}
=\case{i}/2\{Q^{(2)},Q^{(2)*}\}_{\rm DB}
\end{equation}
The combination $Q=(Q^{(1)}+iQ^{(2)})/2$ can also be used, but in this
case
\begin{equation}
i\{Q,Q^*\}_{\rm DB}=H+E_T\label{qq}
\end{equation}
where
\begin{equation}
E_T=2\int g_{ij}(J\partial_z\phi)^i(\partial_{\bar z}\phi)^jd^2z.
\end{equation}
The integral is a constant for fields $\phi$ in the same topological
class. We can still use (\ref{qq}) to generate the quantum
Hamiltonian.

The background field expansion used in section 3 can be used again here
to expand the supercharge $Q$. The bosonic fluctuations about the
background $\phi_0$ are described by a vector $\xi$ and a fluctuation
operator $\Delta_f$. The fermion fields are simply parallel propagated
to $\phi_0$ and decomposed into $\psi=\lambda+\nu^*$, where $\lambda$
and $\nu$ are holomorphic, i.e.  $J\lambda=i\lambda$.

Since the complex structure $J$ commutes with the fluctuation operator,
we can choose holomorphic eigenmodes $u_\alpha$ and use these as a
basis ${\bf e}_\alpha$. The supercharge for this theory is given by
\begin{equation}
Q=\int(\lambda^i\pi_i-2g_{ij}\nu^i\partial_z\phi^j)d^2z
\end{equation}
When the eingenmode expansions are used, we recover the expression
(\ref{sq}) used in section 4,
\begin{equation}
Q=\lambda^\alpha\pi_\alpha-\nu^\alpha W_\alpha(\xi)
\end{equation}
However, now we have an explicit formula for the superpotential
\begin{equation}
W_\alpha=2(u_\alpha,\exp(\nabla_\xi){\bf e}_z)
\end{equation}
in terms of the product (\ref{product}). The series expansion for
$W_{\alpha}$ in the coordinate $z^I$ is obtained using equation
(\ref{serie}). The linear term is,
\begin{equation}
W_{\alpha \bar J}=2(u_\alpha,D_zu_J^*).
\end{equation}
where
\begin{equation}
D_zu_I=\partial_zu_I{}^i+(\partial_z\phi_0^j)\Gamma^i{}_{jk}u_I{}^k
\end{equation}
According to equation (\ref{bmass}), this implies that the fluctuation
operator is
\begin{equation}
\Delta_fu_I=-4D_zD_{\bar z}u_I
\end{equation}
and also that the fermion mass matrix is diagonalised by choosing a
basis
\begin{equation}
{\bf e}_{I'}=2\omega_I{}^{-1}D_{\bar z}u_I.
\end{equation}
These vectors actually form a complete basis, that is they form a
normalisable set of eigenvectors of the (positive) operator $D_{\bar
z}D_z$. The only massless fermions are the $\lambda^a$ and these remain
as fermions on the moduli space. This is precisely the situation
considered in the previous section. The reduced theory is therefore
identical to the truncated theory at leading order.

\section{CONCLUSION}

The reduction of a classical dynamical system generally involves
eliminating the internal forces and introducing generalised
coordinates. In quantum theory, the uncertainty principal implies that
the internal coordinates can never be frozen, but the reduction can be
performed approximately when the internal degrees of freedom remain
close to their ground state.

We have considered the adiabatic reduction of various quantum systems
onto a moduli space of collective coordinates. The results for a
quantum mechanical system are quite general and depend on geometrical
properties of the moduli space as well as the frequencies of the
internal modes. The reduced Hamiltonian to leading order can be found
at the end of section 1.

The moduli space of sigma-model solitons has been quantised by taking
the continuum limit of the quantum mechanical system. In the Bosonic
case, the terms in the reduced Hamiltonian operator depend on the
eigenfunctions of a fluctuation operator $\Delta_f$. The spectrum of
the fluctuation operator is continuous and the eigenvalue sums of the
quantum mechanical system have to be replaced by integrals and
regularised.

We could have attempted to quantise the sigma-model moduli space with
the instanton techniques used in quantum field theory. In this
approach, based on a path integral, the field is replaced by collective
coordinates $x^a$ and field fluctuations $\xi$. The fluctuations are
integrated out, leaving a path integral over the coordinates $x^a$ and
a Jacobian factor. The classical action $S$ gets replaced by
\begin{equation}
W[x^a]=S[x^a]+\case1/2\log\det(\Delta_f-\partial_t^2)'+\dots
\end{equation}
where the prime indicates omission of any zero eigenvalues and the dots
denote contributions from higher loop Feynman diagrams. An adiabatic
reduction can now be obtained by taking the time derivatives to be
small and evaluating some of the higher loop terms. In practice, we
have found this approach less practical and not as rigourous as the
Hamiltonian one presented earlier.

The bosonic sigma model considered in section 3 can be generalised in
various ways. An important possibility would be to include potentials
of the form
\begin{equation}
\Phi=\int W(\phi)d^2x.
\end{equation}
The changes to the results brought about by introducing potentials are
confined to the terms in equation (\ref{result}) involving $V_{IJK}$
and $V_{IJKL}$, which depend on functional derivatives of $\Phi$.

The supersymetric sigma model is tightly constrained and reduces to the
Hodge Laplacian acting on antiholomorphic $p$-forms, confirming a
result that has been used previously with strong support but not
derived rigourously \cite{witten,alfaro,gauntlett}. It is here,
especially, that higher order terms in the adiabatic expansion might be
of interest.


\appendix

\section{FACTOR ORDERING}

When constructing a quantum theory the ordering of operators is an
important issue. A minimum requirement for choosing the operator
ordering is to try to retain as much of the symmetry of the classical
system as possible. We shall consider symmetry under general coordinate
transformations, supersymmetry and the symmetry of a Riemannian
Symmetric space.

Our first example is classical particle mechanics in curved space,
described by the Hamiltonian
\begin{equation}
H=\case1/2g^{\alpha\beta}p_\alpha p_\beta.
\end{equation}
If we introduce a Hilbert space of wave functions $\psi(x)$ and use the
metric to construct an inner product on the Hilbert space, then the
Hamiltonian operator
\begin{equation}
H=-\case1/2|g|^{-1/2}\partial_\alpha|g|^{1/2}g^{\alpha\beta}\partial_\beta.
\end{equation}
is both covariant and consistent with the basic principals of quantum
mechanics. However, many other possibilities exist, since we are free
to add terms such as
\begin{equation}
i\xi R^\alpha{}_\beta[x^\alpha,p_\beta]
\end{equation}
which are of order $\hbar$ or higher and vanish in the classical limit.
If we assume that the Hamiltonian operator is a covariant second order
operator, then the form of this operator must be \cite{dewitt}
\begin{equation}
H=-\case1/2(\nabla+A)^2+X
\end{equation}
However, the vector field $A$ and scalar field $X$ cannot be infered
from the classical Hamiltonian.

The quantum theory is far more restrictive if the space is a Riemannian
Symmetric space. A Riemannian Symmetric space is defined by its
geodesic symmetries. The geodesic symmetries about a point leave the
point unchanged and map each of the geodesics through the point into
itself. A Riemannian Symmetric space has these symmetries at every
point and covariant derivatives of the curvature, which are odd under
the reflection symmetry, must vanish \cite{besse}. In this case, both
$\nabla X$ and $A$, which are also odd under the geodesic symmetry,
must vanish. The Hamiltonian on a Riemannian symmetric space is
therefore given by the Laplacian and a constant.

Our second example is the supersymmetric particle mechanics on a Kahler
manifold, originally discussed by Witten \cite{witten}. The coordinates
$z^a$ and momenta $\pi_a$ are supplimented by a set of complex fermions
$\lambda^a$. The commutation relations are
\begin{eqnarray}
\{\lambda^a,\lambda^{\bar b}\}&=&g^{a\bar b}\\
\lbrack\pi_a,z^b]&=&-i\delta_a{}^b\\
\lbrack\pi_a,\lambda^b]&=&i\Gamma^b{}_{ca}\lambda^c\\
\lbrack\pi_a,\pi_{\bar b}]&=&iR_{a\bar bd\bar c}\lambda^{\bar
c}\lambda^d
\end{eqnarray}
and their hermitian conjugates, where $g^{a\bar b}$ is the inverse
metric, $\Gamma^b{}_{ca}\lambda^c$ the connection components and
$R_{a\bar bd\bar c}$ the Riemannian curvature. The supersymmetry is
generated by a supercharge
\begin{equation}
Q=\lambda^a\pi_a.\label{supch}
\end{equation}
The ordering of the operators in the last commutation relation and the
supercharge have so far been chosen arbitrarily.

The commutation relations can be represented by differential operators
acting on the Hilbert space of functions $\Psi(z^a,z^{\bar
a},\lambda^{\bar a})$, with
\begin{eqnarray}
\pi_a&=&-i\partial_a\\
\pi_{\bar a}&=&-i(\partial_{\bar a}-
\Gamma_{c\bar b\bar a}\lambda^{\bar b}\lambda^c)
\end{eqnarray}
where
\begin{equation}
\lambda^a=g^{a\bar b}{\partial\over\partial\lambda^{\bar b}}.
\end{equation}
In this representation all of the operators, including $Q$, act
covariantly on the functions $\Psi$. The covariance is explicit if we
regard $\Psi$ as an element of the exterior algebra over a basis of
forms $\lambda^{\bar a}$. In this case $\pi\lambda^{\bar
a}=-i\nabla\lambda^{\bar a}$, where $\nabla$ is the covariant
derivative, and $Q=i\bar\partial^\dagger$, where $\bar\partial$ is the
antiholomorphic exterior derivative.

The Hamiltonian is obtained from $Q$ by the relation $H=\{Q,Q^*\}$,
which by the use of equation (\ref{supch}) and the commutation
relations gives
\begin{equation}
H=-\case1/2\nabla^2+\case1/2R_{\bar a b}\lambda^{\bar a}\lambda^b
\end{equation}
where $R_{\bar a b}$ is the Ricci tensor.

De Alfaro et al. \cite{alfaro} have pointed out that alternative
orderings of the two operators in the supercharge (\ref{supch}) would
make the representation non-covariant. However, there do exist
alternative expressions for the supercharge which have the same
classical limit,
\begin{equation}
Q=\lambda^a\pi_a-i\lambda^aW_a(z)+\lambda^{\bar a}U_{\bar a}(z)
\end{equation}
The condition $Q^2=0$ is satisfied if $\partial W=0$ and
$\bar\partial^\dagger U=0$. The simplest way to satisfy these
conditions would be to make $W$ a total derivative,
\begin{equation}
W_a=i\xi \partial_a(R^b{}_c[x^c,p_b])
\end{equation}
which gives additional contributions to the Hamiltonian of order
$\hbar^3$. Other contributions to the Hamitonian can exist if the
manifold has non-trivial Homology.

\section{GEOMETRICAL FORMULAE}

This appendix reviews some formulae used to reduce the Riemannian
connection $\nabla$ onto a submanifold ${\cal M}$. We construct a basis
of 1-forms $({\bf n}^I,{\bf e}^a)$, with dual basis $({\bf n}_I,{\bf
e}_a)$, made up to include the normal forms ${\bf n}^I$ and vectors
${\bf e}^a$ tangential to ${\cal M}$. The metric is then
\begin{equation}
{\bf g}=\sigma_{ab}{\bf e}^a\otimes{\bf e}^b
+\eta_{IJ}{\bf n}^I\otimes{\bf n}^J
\end{equation}
where $\sigma$ is the metric on ${\cal M}$. Any vector ${\bf X}$ can be
decomposed into a tangential and normal parts,
\begin{eqnarray}
{\bf X}_\parallel={\bf e}^a({\bf X}){\bf e}_a\\
{\bf X}_\perp={\bf n}^I({\bf X}){\bf n}_I
\end{eqnarray}

The extrinsic curvature ${\bf k}^I$ and extrinsic torsion tensor ${\bf
a}_I{}^J$ are defined by
\begin{eqnarray}
k^I{}_{ab}&=&(\nabla_a{\bf n}^I)_b\\
a_I{}^J{}_a&=&(\nabla_a{\bf n}^J)_I
\end{eqnarray}
The extrinsic curvature measures the expansion of the normal forms and
it is always symmetric in the surface indices. The extrinsic torsion
measures changes in the normals along lines drawn in the surface. The
extrinsic torsion is antisymmetric if the normals are of unit length.

We will define the tangential covariant derivative by its action on the
basis vectors,
\begin{equation}
{}^\parallel\nabla_{\bf X}{\bf e}_a=
(\nabla_{X_\parallel}{\bf e}_a)_\parallel,\qquad
{}^\parallel\nabla_{\bf X}{\bf n}_I=(\nabla_{X_\parallel}{\bf
n}_I)_\perp
\end{equation}
With this definition the derivative is a metric connection for both
$\sigma$ and $\eta$, with torsion related to ${\bf a}_I{}^J$. We make
repeated use of the following expression for the derivative of a normal
directed tensor in the body of the text,
\begin{equation}
{}^\parallel\nabla_a\omega_{IJ}=\partial_a\omega_{IJ}
+a_I{}^K{}_a\omega_{KJ}+a_J{}^K{}_a\omega_{IK}\label{grado}
\end{equation}
The curvature tensor of ${}^\parallel\nabla$ will be denoted by
${}^\parallel R$.

The reduction of the curvature tensor is a standard proceedure, and has
been described in this notation in reference \cite{grant}. The most
useful formula is Gauss equation,
\begin{equation}
R_{abcd}={}^\parallel R_{abcd}-\eta_{IJ}k^I{}_{ac}k^J{}_{bd}
+\eta_{IJ}k^I{}_{ad}k^J{}_{bc}\label{gauss}
\end{equation}
In the case where ${\cal M}$ is a complex submanifold of a K\"ahler
manifold, the trace $k^I$ of the extrinsic curvature tensor vanishes
\cite{lawson}.

\section{ADIABATIC APPROXIMATION SCHEMES}

This appendix shows how the hamiltonian operator and other operators
can be reduced by using almost degenerate perturbation theory
\cite{sakuri}. The method is equivalent to the adiabatic, or
Born-Oppenheimer, approximation scheme for the energy eigenstates.

The Hamiltonian is first separated into an unperturbed part and a
perturbation
\begin{equation}
H=H_0+H_1.
\end{equation}
The eigenstates of $H_0$ are then grouped into eigenspaces that cover a
narrow range of eigenvalues. A projection operator $P_0$ projects onto
the lowest of these eigenspaces and $P_1=1-P_0$. The equation for the
eigenstate with energy $E$ is standard (see \cite{sakuri} for example),
\begin{equation}
H_R\,P_0\Psi=E\,P_0\Psi
\end{equation}
where
\begin{equation}
H_R=E_0+P_0H_1P_0+P_0H_1P_1(E-H_0-H_1)^{-1}P_1H_1P_0\label{hamr}
\end{equation}
The leading order terms give
\begin{equation}
H_R=E_0+P_0H_1P_0+P_0H_1P_1(E_0-H_0)^{-1}P_1H_1P_0+\dots\label{hrexp}
\end{equation}

For an adiabatic approximation scheme, we look for an expansion of the
wave function in the form
\begin{equation}
\Psi(x)=\sum_n\psi(x^a)f_n(x^I,x^a)
\end{equation}
where the wave function depends strongly on coordinates $x^I$ and
weakly on coordinates $x^a$. We take an unperturbed Hamiltonian $H_0$
which commutes with the $x^a$ and define the functions $f_n$ by
\begin{equation}
H_0\,f_n=E_n(x^a)\,f_n.
\end{equation}
These can be normalised with respect to a reduced product
\begin{equation}
\langle f,g\rangle_\perp=\int f^*(x^a,x^I)g(x^a,x^I)d\mu(x^I).
\end{equation}
with measure $d\mu(x^I)$ chosen to make $H_0$ self-adjoint.

The projection operator $P_0$ is given by
\begin{equation}
P_0\Psi=\psi_0(x^a)
\end{equation}
Inserting complete sets of states into equation (\ref{hrexp}) gives the
equation for $\psi_0$,
\begin{equation}
H_R\psi_0=E\psi_0
\end{equation}
where the reduced Hamiltonian operator has the perturbative expansion
\begin{equation}
H_R=E_0+\langle H_1\rangle_{00}+
\sum_m\langle H_1\rangle_{0m}(E_0-E_m)^{-1}\langle H_1\rangle_{m0}
+\dots
\end{equation}
The matrix elements
\begin{equation}
\langle H_1\rangle_{mn}=\langle f_m,H_1f_n\rangle_\perp
\end{equation}
are operators of the reduced theory. They also act on the $E_n$.

Other operators are reduced in a similar way. Given an operator $Q$, we
follow (\ref{hamr}) and define
\begin{equation}
Q_R=P_0QP_0+P_0QP_1(E-H)^{-1}P_1HP_0.\label{qr}
\end{equation}
In the adiabatic approximation,
\begin{equation}
Q_R=\langle Q\rangle_{00}+
\sum_n\langle Q\rangle_{0n}(E_0-E_n)^{-1}
\langle H_1\rangle_{n0}+\dots
\end{equation}
is an operator that acts on the states $\psi_0$.

The reduced operator satisfies
\begin{equation}
Q_RP_0\Psi=P_0Q\Psi
\end{equation}
where $\Psi$ is any energy eigenstate with eigenvalue $E$. This
reduction is not generally a homomorphism of the operator algebras,
i.e. $(QQ')_R\ne Q_RQ'_R$ in general, but if $Q$ and $Q'$ are
symmetries of the Hamiltonian, then it follows from (\ref{qr}) that
\begin{equation}
(QQ')_R=Q_RQ'_R
\end{equation}
In particular, if $Q$ is a supersymmetry operator with $H=\{Q,Q^*\}$,
then
\begin{equation}
H_R=\{Q_R,Q^*_R\}
\end{equation}
This simplifies the problem of reducing the Hamiltonian operator.

\end{document}